\documentclass[
	reprint,
	superscriptaddress,
	showkeys,
	aps,
	pra,
	longbibliography,
	nofootinbib
]{revtex4-2}

\usepackage[utf8]{inputenc}

\usepackage{xcolor}
\definecolor{blau}{cmyk}{1.0,0.2,0.0,0.4}
\definecolor{rot}{cmyk}{0.04,1.0,0.8,0.07}
\definecolor{purple}{cmyk}{0.08,1.0,0.3,0.36}

\usepackage[
	colorlinks,
	pdfpagelabels,
	breaklinks,
	pdfstartview=FitH,
	bookmarksopen=true,
	bookmarksnumbered=true,
	bookmarksopenlevel=2,
	plainpages=false,
	hypertexnames=false,
	citecolor=rot,
	linkcolor=blau,
	urlcolor=purple,
	pdftitle={Regulator Optimization},
	pdfauthor={Niklas Zorbach, Jonas Stoll, Jens Braun},
	pdfkeywords={regulator optimization, principle of strongest singularity, symmetry breaking/restoration, FRG, numerical fluid dynamics}
]{hyperref}

\usepackage{bm,dsfont}
\usepackage{tensor}
\usepackage{amsmath,amssymb,amsthm}
\usepackage{physics}
\usepackage{upgreek}
\usepackage[bb=boondox]{mathalfa}
\usepackage{mathtools}

\usepackage{bookmark}
\usepackage{graphicx}
\usepackage[caption=false]{subfig}
\usepackage{dcolumn}

\usepackage[acronym]{glossaries-extra}
\glssetcategoryattribute{acronym}{nohyperfirst}{true}
\setabbreviationstyle[acronym]{long-short}

\newacronym{UV}{UV}{ultraviolet}
\newacronym{IR}{IR}{infrared}
\newacronym{POSS}{POSS}{Principle of Strongest Singularity}
\newacronym{PMS}{PMS}{Principle of Minimum Sensitivity}
\newacronym{RG}{RG}{Renormalization Group}
\newacronym{LPA}{LPA}{local potential approximation}
\newacronym{KT}{KT}{Kurganov-Tadmor}
\newacronym{SD}{SD}{strong divergent}
\newacronym{NSL}{NSL}{naively smeared Litim}
\newacronym{OSC}{OSC}{oscillatory}
\newacronym{SL}{SL}{smooth Litim-like}
\newacronym{CSS}{CSS}{compactly supported smooth}
\allowdisplaybreaks

\newcommand\numberthis{\addtocounter{equation}{1}\tag{\theequation}}
\newcommand{\orcid}[1]{\href{https://orcid.org/#1}{\includegraphics[height=1.9ex,width=1.9ex]{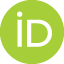}}}

\def\d{\,\mathrm{d}}

\usepackage[capitalise,sort,compress]{cleveref}
\Crefname{section}{Sec.}{Sections}
\Crefname{table}{Tab.}{Tables}

\begin{document}

\title{Optimization and Stabilization of Functional Renormalization Group Flows}

\author{Niklas Zorbach~\orcid{0000-0002-8434-5641}}
\email{niklas.zorbach@tu-darmstadt.de}
\affiliation{
	Technische Universität Darmstadt, Department of Physics, Institut für Kernphysik, Theoriezentrum,\\
	Schlossgartenstraße 2, D-64289 Darmstadt, Germany
}

\author{Jonas Stoll~\orcid{0000-0001-8204-9876}}
\email{jonas.stoll@tu-darmstadt.de}
\affiliation{
	Technische Universität Darmstadt, Department of Physics, Institut für Kernphysik, Theoriezentrum,\\
	Schlossgartenstraße 2, D-64289 Darmstadt, Germany
}

\author{Jens Braun~\orcid{0000-0003-4655-9072}}
\email{jens.braun@tu-darmstadt.de}
\affiliation{
	Technische Universität Darmstadt, Department of Physics, Institut für Kernphysik, Theoriezentrum,\\
	Schlossgartenstraße 2, D-64289 Darmstadt, Germany
}
\affiliation{ExtreMe Matter Institute EMMI, GSI, Planckstra{\ss}e 1, D-64291 Darmstadt, Germany}

\begin{abstract}
	We revisit optimization of functional renormalization group flows by analyzing regularized loop integrals. This leads us to a principle, the {\it Principle of Strongest Singularity}, and a corresponding order relation which allows to order existing regularization schemes with respect to the stability of renormalization group flows. Moreover, the order relation can be used to construct new regulators in a systematic fashion. For studies of critical behavior, which require to follow renormalization group flows down to the deep infrared regime, such new regulators may turn out to be particularly useful. The general application of this principle is demonstrated with the aid of a scalar field theory which is solved over a wide range of scales with novel methods borrowed from numerical fluid dynamics.
\end{abstract}

\maketitle

\section{Introduction}\label{sec:intro}
In quantum field theory, the computation of observables in general requires to introduce a regularization. Of course, any physical observable~${\mathcal O}$ should be independent of the chosen regularization scheme~$R$, i.e.,
\begin{align*}
	\numberthis
	\frac{\delta {\mathcal O}}{\delta R} =0\,.
\end{align*}
Strictly speaking, however, this independence of the regularization scheme only holds for the exact result for~$\mathcal O$. In practice, the computation of observables requires a truncation of the quantum effective action~$\Gamma$ of some kind. In a truncated calculation, it then appears natural to search for an optimal regularization scheme which brings the results as close as possible to the {\it a priori} unknown exact results for the observables under consideration.

The optimization of the computation of the effective action and derived observables with respect to the regularization scheme has been discussed intensely over decades now and its origin may be dated back to the seminal work by Stevenson on optimized perturbation theory and the introduction of the \gls{PMS}~\cite{Stevenson:1981vj}. Loosely speaking, this principle relies on the assumption that the optimal scheme defines a stationary point of an observable or a given set of observables under the variation of the regularization scheme.

In the context of the functional \gls{RG} framework, which also underlies our present analysis, the starting point for the computation of any observable is the Wetterich equation~\cite{Wetterich:1992yh}:
\begin{align*}
	\numberthis\label{eq:WetterichEq}
	\partial_t \Gamma_k[\Phi] = \frac{1}{2} \mathrm{STr}\left[\left(\Gamma_k^{(2)}[\Phi] + R_k\right)^{-1} \cdot \partial_t R_k\right] \,.
\end{align*}
Here, $\Gamma_k[\Phi]$ denotes the scale-dependent (i.e., coarse-grained) effective average action. The supertrace arises since the field~$\Phi$ of a given theory may in general contain both bosonic and fermionic degrees of freedom and therefore provides a minus sign in the fermionic subspace. The regulator function $R_k$ depends on the \gls{IR} cutoff scale $k$ and defines the regularization scheme. The \gls{RG} time is related to the scale~$k$ via $t=-\ln(k/\Lambda)$ with~$\Lambda$ being the \gls{UV} scale.

Optimization of \gls{RG} flows has already been discussed early on by means of a \gls{PMS}-like criterion in Ref.~\cite{Liao:1999sh}. In the context of the functional \gls{RG} approach, \gls{PMS} and the associated optimization have been considered and employed in computations of critical exponents of scalar field theories, see, e.g., Refs.~\cite{Canet:2002gs,Canet:2003qd}. For a more recent discussion, including computations of universal amplitude ratios, we refer to Refs.~\cite{Balog:2019rrg,DePolsi:2020pjk,DePolsi:2021cmi,DePolsi:2022wyb}. A first detailed and general discussion of optimization of functional \gls{RG} flows has been given by Litim in a series of ground-breaking articles~\cite{Litim:2000ci,Litim:2001up,Litim:2001fd}. Litim's optimization criterion is based on a maximization of the gap in the regularized two-point function in the \gls{RG} flow and encompasses \gls{PMS} as a special case. In fact, the optimization introduced by Litim is more general and the underlying criterion eventually leads to an improved convergence of the \gls{RG} flow, see also Ref.~\cite{Pawlowski:2005xe}. Studies of critical exponents based on Litim's optimization criterion can be found in Refs.~\cite{Litim:2002cf,Litim:2003kf,Litim:2007jb,Litim:2010tt}.

In a seminal analysis of optimization of functional \gls{RG} flows~\cite{Pawlowski:2005xe}, Pawlowski then showed that, loosely speaking, Litim's criterion for the construction of an optimized regulator based on the maximization of the gap in the propagator has to be supplemented by the minimization of the full (regularized) two-point function entering the flow equation~\eqref{eq:WetterichEq} to obtain optimized regulators for truncations at any given order~$n$ of the derivative expansion.\footnote{This minimization requires to compute the norm of the full propagator on Sobolev spaces $H_n$ with $n\in {\mathbb N}$.} 

Phenomenologically speaking, the optimization discussed by Litim and Pawlowski aims at a minimization of the deviation of the values of observables computed with a given truncation from their exact values. For example, this approach to optimization of \gls{RG} flows can be employed to improve computations of critical exponents but also underlies other type of studies, such as an early study of the deconfinement phase transition in Yang-Mills theory~\cite{Braun:2007bx}. Note that a prescription for the construction of such optimized regulators based on a suitable definition of the length of \gls{RG} trajectories can be found in Ref.~\cite{Pawlowski:2015mlf}. Last but not least, we add that it has more recently also been proposed to optimize functional \gls{RG} flows by employing constraints originating from conformal invariance~\cite{Balog:2020fyt,Delamotte:2024xhn}. 

The optimization criterion discussed in our present work aims at an improvement of the stability of functional \gls{RG} flows in the presence of spontaneous symmetry breaking and emerges from an analysis of \gls{RG} flows computed with techniques borrowed from the field of numerical fluid dynamics, which provide an intuitive picture of the dynamics of \gls{RG} flows in terms of advection-diffusion equations~\cite{Grossi:2019urj,Grossi:2021ksl,Koenigstein:2021syz,Koenigstein:2021rxj,Koenigstein:2023wso,Steil:2021cbu}.

In case of theories featuring spontaneous symmetry breaking, convexity of the effective action implies the emergence of a flat region in field space in the \gls{RG} flow which may eventually cause numerical precision problems, i.e., breakdowns of \gls{RG} flows at (comparatively large) finite \gls{RG} scales. In general, the precise value of this breakdown scale depends on the chosen regulator. In the present work, we show that the stability of the \gls{RG} flow in such a situation is directly related to the strength of the singularities of the {\it regularized} loop integrals in the  regime of negative mass parameters. This naturally results in the definition of an order relation on the space of regulators. To illustrate this, we employ the so-called \gls{LPA}, the zeroth-order of the derivative expansion of the effective action. However, our line of arguments is not bound to this approximation but is more general, as we shall discuss right at the very beginning in \cref{sec:poss}. There, we introduce the principle underlying our optimization, the \gls{POSS}, as well as the associated order relation. In \cref{sec:construction-of-regulators}, we employ this principle to discuss regulator functions and also to construct new regulators as concrete examples for the application of this principle. In \cref{sec:numerical-stability-tests}, we then numerically test the stability of \gls{RG} flows obtained from these new regulators and also from other existing regulators by computing the \gls{RG} flow of the effective potential of a simple scalar field theory in \gls{LPA}. Our conclusions can be found in \cref{sec:conc}. 

\section{Principle of strongest singularity}\label{sec:poss}
\subsection{General Analysis}
In the derivation of the Wetterich equation \eqref{eq:WetterichEq} one tacitly assumes that the scale-dependent full propagator, 
\begin{align*}
G_k[\Phi] = \left(\Gamma^{(2)}_k[\Phi] + R_k\right)^{-1}\,,
\numberthis\label{eq:Gdef}
\end{align*}
exists for all $k > 0$. From an analysis of the general form of the Wetterich equation, it follows that this assumption is justified. Indeed, the \gls{RG} flow generated by the Wetterich equation is such that it tends to push the inverse propagator away from potentially existing points with eigenvalue zero. This becomes more apparent by realizing that the Wetterich equation \eqref{eq:WetterichEq} is of the form 
\begin{align*}
	\numberthis\label{eq:self-healing}
	\partial_t \Gamma_k \propto -\det(\Gamma^{(2)}_k[\Phi] + R_k)^{-1}\,.
\end{align*}
Note that we have defined the \gls{RG} time to be a positive quantity, $t=-\ln(k/\Lambda) > 0$, and that $\det G_k>0$ since the eigenvalues of the operator in \cref{eq:Gdef} are positive.
Thus, if the \gls{RG} flow approaches a point associated with zero eigenvalues, then the determinant of the inverse scale-dependent propagator becomes smaller and smaller. This may be considered as a``self-healing" property of the Wetterich equation since it eventually prevents the \gls{RG} flow from reaching a point with zero eigenvalues. Note that the appearance of zero eigenvalues in the eigenvalue spectrum of the inverse propagator is intimately connected with the formation of a convex effective action, see also Ref.~\cite{Litim:2006nn}. In particular, we encounter this scenario in theories which develop a nontrivial minimum of the effective action in the \gls{RG} flow that survives in the \gls{IR} limit, i.e., theories featuring spontaneous symmetry breaking. The focus of the present work is exactly on this class of theories. In any case, since the inverse scale-dependent propagator depends on the regulator, it is clear that the regulator choice affects the ``strength'' of the singular behavior close to the zero eigenvalues of the inverse propagator. Below, this observation leads us to the \acrlong{POSS}. The underlying idea is that regulators, which generate \gls{RG} flows with {\it optimized} self-healing properties, eventually yield more stable \gls{RG} flows. Based on this principle, we shall also introduce an order relation which allows to compare regulators in a meaningful manner.

Before we can quantitatively compare different regulator schemes, we have to introduce a {\it comparability condition}. To this end, we consider a theory with spontaneous symmetry breaking and assume that $\Gamma_k$ is a {\it generic} reference solution of the Wetterich equation~\eqref{eq:WetterichEq} obtained with a given reference regulator $R_k$. Then, there exists at least one field configuration $\Phi_0$ where the determinant of the inverse scale-dependent propagator vanishes in the \gls{IR} limit, $k\to 0$:\footnote{Since the effective action becomes convex in the \gls{IR} limit~\cite{Litim:2006nn}, there in general exists a set of values of $\Phi$ for which the determinant in \cref{eq:vanishing_determinant} vanishes.}\footnote{The factor $1/k^2$ in \cref{eq:vanishing_determinant} is necessary to render the determinant dimensionless at least for bosonic fields.}
\begin{align*}
	\numberthis\label{eq:vanishing_determinant}
	\lim_{k\to 0} \det(\frac{1}{k^2}(\Gamma^{(2)}_k[\Phi_0] + R_k)) = 0 \, ,
\end{align*}
i.e., the full inverse propagator can not be inverted at $\Phi_0$ in this limit. Now let $\tilde R_k$ be a regulator which differs from~$R_k$. We then say that the regulator $\tilde R_k$ is {\it comparable} to $R_k$, if
\begin{align*}
	\numberthis\label{eq:comparability_condition0}
	\lim_{k\to 0} \det(\frac{1}{k^2}(\Gamma^{(2)}_k[\Phi_0] + \tilde R_k)) = 0 \,
\end{align*}
for the same field configuration~$\Phi_0$, i.e., we do {\it not} have
\begin{align*}
	\numberthis\label{eq:comparability_condition1}
	\lim_{k\to 0} \det(\frac{1}{k^2}(\Gamma^{(2)}_k[\Phi_0] + \tilde R_k)) &> 0
\end{align*}
or 
\begin{align*}
	\numberthis\label{eq:comparability_condition2}
	\lim_{k\to k_c > 0} \det(\frac{1}{k^2}(\Gamma^{(2)}_k[\Phi_0] + \tilde R_k)) &= 0 
\end{align*}
for a nonzero $k_c$. We emphasize that $\Gamma_k$ is a solution of the Wetterich equation with respect to $R_k$ not $\tilde R_k$. In~\cref{eq:comparability_condition0}, we simply state that a combination of a given $\Gamma_k$ obtained from a given regulator with another regulator $\tilde R_k$ does not alter the limiting behavior \eqref{eq:vanishing_determinant}.

With this comparability condition at hand, we can now compare how different regulators behave with respect to a given reference solution $\Gamma_k$ in the \gls{IR} limit. A first guess for an order relation for regulators on the space of all comparable regulators would be to compare how ``fast'' the determinant at a given field configuration~$\Phi_0$ vanishes. Loosely speaking, a regulator yielding a rapidly vanishing determinant is associated with a stronger singular behavior close to~$\Phi_0$ on the right-hand side of the Wetterich equation, see \cref{eq:self-healing}. Consequently, such a stronger singular behavior improves the self-healing property of the \gls{RG} flow and hence its stability. However, this reasoning ignores the exact form of the Wetterich equation. A meaningful comparison is obtained by considering the full right-hand side of the Wetterich equation in the \gls{IR} limit. Assume $R^A$ and $R^B$ are comparable to a given $R_k$ according to our definition above, then we can define an order relation as follows:
\begin{align*}
	\numberthis\label{eq:general_POSS}
	\mathcal{R}(R^A, R^B) = \lim_{k \to 0} \frac{\mathrm{STr}\left((\Gamma^{(2)}_k[\Phi_0] + R^A_k)^{-1}\cdot \partial_t R^A_k\right)}{\mathrm{STr}\left((\Gamma^{(2)}_k[\Phi_0] + R^B_k)^{-1}\cdot \partial_t R^B_k\right)} \, .
\end{align*}
Recall again that $\Gamma_k$ is a reference solution of the Wetterich equation obtained with a given reference regulator~$R_k$. If $\mathcal{R}(R^A, R^B) > 1$ the singularity of $R^A$ is stronger than of $R^B$ and, in turn, yields a more stable \gls{RG} flow. We shall therefore write $R^A > R^B$ (i.e., {\it $R^A$ is more stable than $R^B$}) if $\mathcal{R}(R^A, R^B) > 1$ and $R^A < R^B$ (i.e., {\it $R^A$ is less stable than $R^B$}) if $\mathcal{R}(R^A, R^B) < 1$. Hence, $\mathcal{R}$ defines an order relation which is reflexive, transitive and strongly connected.\footnote{In general, this order relation depends on $\Phi_0$ and on the reference \gls{RG} flow $\Gamma_k$ associated with a corresponding reference regulator~$R_k$. By using a different reference flow or a different field configuration $\Phi_0$, we could in principle obtain a different order relation.}

\subsection{Local potential approximation}\label{sec:poss_LPA}
In this subsection, we would like to {\it illustrate} the application of the \acrlong{POSS} and the associated order relation \eqref{eq:general_POSS} with the aid of a concrete example. To be specific, we shall consider a scalar field theory in $d$ spacetime dimensions in \gls{LPA}. The classical action is given by
\begin{align*}
	\numberthis
	S[\phi] = \int {\rm d}^dx\left\{ \frac{1}{2}(\partial_{\mu}\phi)^2 + \frac{1}{2}m^2\phi^2 + \frac{1}{4}\lambda\phi^4\right\}\,,
\end{align*}
where~$\phi$ is a real-valued scalar field. This action serves as the initial condition for our ansatz for the scale-dependent effective action~$\Gamma_k$:
\begin{align*}
	\numberthis
	\Gamma_k[\phi] = \int {\rm d}^dx\left\{ \frac{1}{2}(\partial_{\mu}\phi)^2 + U_k(\phi)\right\}\,.
\end{align*}
For~$k\to\Lambda$, we have $\Gamma_k \to S$. By plugging this ansatz into the Wetterich equation, see \cref{eq:WetterichEq} and evaluating it then on a constant field configuration $\phi$, we obtain the flow equation for the scale-dependent effective potential~$U_k$. For concreteness, we shall restrict ourselves to regulators of the form $R_{k}(p, q) \propto R_k(p^2) \delta^{(d)}(p+q)$, where $R_k(p^2) = p^2 \, r(p^2/k^2)$ can be expressed through a dimensionless regulator shape function $r(y)$ with~$y=p^2/k^2$ and~$p^2=p_0^2+ p_1^2 +\dots +p_{d-1}^2$. 

An illustrative analysis of the stability of \gls{RG} flows for this model based on \gls{POSS} can be found in \cref{sec:numerical-stability-tests}. In the following, however, we shall first focus on a general discussion of \gls{POSS} in \gls{LPA} and derive the order relation for regulators in this approximation. For this discussion, it is convenient to introduce the following auxiliary quantity:
\begin{align*}
	\numberthis\label{eq:P2}
	P^2(y)=y(1+r(y))\,.
\end{align*}
From this object, we can construct the so-called threshold functions which correspond to one-particle irreducible one-loop diagrams and appear as the fundamental building blocks in the flow equations of, e.g., the effective potential and the couplings. The regularization-scheme dependence of the flow equations is solely encoded in these functions. In a calculation of the effective potential in \gls{LPA}, they are of the form
\begin{align*}
	\numberthis\label{eq:generic-threshold-function}
	l^{(d)}_m(\omega) = \int_0^{\infty} \d y\, I^{(d)}_m(\omega, r(y), r'(y), y)
\end{align*}
with
\begin{align*}
	\numberthis\label{eq:generic-threshold-function-integrand}
	I^{(d)}_m(\omega, r(y), r'(y), y) =  -y^{\frac{d}{2}+1} \frac{\partial_y r(y)}{(P^2(y) + \omega)^m} \,
\end{align*}
and~$m\geq 1/2$. As such they are functionals of $r(y)$ or equivalently, of $P^2(y)$ because of Eq.~\eqref{eq:P2}. The parameter~$m$ ``measures" the number of internal lines of the threshold function. The parameter~$\omega$ plays the role of an effective (potentially field-dependent) dimensionless mass term. For example, in our scalar field theory, the parameter~$\omega$ is essentially given by the second derivative of the effective potential with respect to the field, i.e., $\omega = \partial^2_\phi U_k(\phi)/k^2$. 

With the threshold functions at hand, the flow equation for the scale dependent effective potential~$U_k$ assumes the following form:\footnote{In terms of concrete models, for example, this flow equation describes the scalar contribution to the \gls{RG} flow of the effective potential in the Gross-Neveu-Yukawa model, see, e.g., Ref.~\cite{Stoll:2021ori}, but it may also be employed to study critical behavior in the Ising model.}
\begin{align*}
	\numberthis\label{eq:example_LPA_flow}
	\partial_t U_k(\phi) = -\frac{1}{2}\frac{\mathrm{surf}(d) k^d}{(2\pi)^d}\, l^{(d)}_{1}(\omega) \,,
\end{align*}
where $\mathrm{surf}(d)$ is the surface of a $d$-dimensional ball with radius one. To be specific, we have $\mathrm{surf}(d)/d=\mathrm{vol}(d)=2\pi^{d/2}/(d\,\Gamma(d/2))$.

We can now apply the comparability condition \eqref{eq:comparability_condition0} from \cref{sec:poss} to \gls{LPA}. To this end, we first have to determine the full inverse regularized propagator evaluated for some reference field configuration. We choose $\Phi_0 = \phi = 0$, being a simple choice for studies of models featuring a nontrivial minimum of the effective action in the \gls{IR} limit: 
\begin{eqnarray}
	&& \Gamma_{k, pq}^{(2)}[0] + R_{k,pq} \\
	&&\qquad  = (p^2 + \partial^2_\phi U_k (0) + p^2 r(p^2/k^2))(2\pi)^d\delta^{(d)}(p+q) \, . \nonumber
\end{eqnarray}
Using the definition~\eqref{eq:P2}, the (dimensionless) determinant reads
\begin{eqnarray}
	&& \det(\frac{1}{k^2}(\Gamma_{k, pq}^{(2)}[0] + R_{k,pq})) \nonumber \\ 
	&&\qquad = \prod_{p_0, \dots, p_{d-1}} (P^2(y) + \partial^2_\phi U_k(0)/k^2) \,. \nonumber
\end{eqnarray} 
Assuming that the potential develops a nontrivial minimum~$\phi_{\mathrm{min}}$ in the \gls{RG} flow, convexity of the effective action in the \gls{IR} limit requires the determinant to vanish at $\phi = 0$ for~$k\to 0$, implying that $\omega_0(k) = \partial^2_\phi U_k(0)/k^2$ approaches~$\omega_{\rm pole}$,
\begin{align*}
	\numberthis\label{eq:omega-pole}
	\omega_{\mathrm{pole}}=-\min_{0\leq y \leq \infty} P^2(y) \,,
\end{align*}
in this limit. 

From \cref{eq:comparability_condition0}, it now follows that a different regulator~$\tilde{R}$ parametrized by the shape function~$\tilde{r}$ is comparable to the regulator~$R$ parametrized by the shape function~$r$, if $\tilde{P}^2(y)$ associated with~$\tilde{r}$ yields the same $\omega_{\mathrm{pole}}$ in \cref{eq:omega-pole}. In addition, from a comparison of \cref{eq:self-healing} with \cref{eq:example_LPA_flow}, we deduce that the threshold functions~\eqref{eq:generic-threshold-function} should diverge in the limit $\omega_0 \to \omega_{\mathrm{pole}}$ but remain finite for $\omega_0 > \omega_{\mathrm{pole}}$. This singularity of the threshold functions essentially governs the ``self-healing'' property of the \gls{RG} flow mentioned in \cref{sec:poss} and controls the approach of the effective potential to a convex function in the \gls{IR} limit. As already indicated above, the existence of this singularity is necessary for the effective potential to become convex in studies where the ground state is governed by spontaneous symmetry breaking, see also Refs.~\cite{Litim:2006nn,Pelaez:2015nsa}.

To summarize, we deduced from the comparability condition \eqref{eq:comparability_condition0} that $\omega_{\mathrm{pole}}$ must assume the same value for any two regulators in order to render them comparable, and as a consequence, the position of the singularity of the threshold functions must also be the same.
In the following we choose $\omega_{\mathrm{pole}} = -1$ without loss of generality and consider only regulators~$R$ which fulfill
\begin{align*}
	\numberthis\label{eq:normalization-condition}
	\min\limits_{0 \leq y \leq \infty} P^2(y) = 1\,.
\end{align*}
Note that this relation defines a {\it normalization condition} in the sense that {\it any} regulator~$R$ of the form~$R_k(p^2)=p^2\, r(p^2/k^2)$ can be {\it normalized} such that it satisfies \cref{eq:normalization-condition} and hence can be rendered comparable. In fact, this can be achieved by exploiting the freedom that we can rescale the unphysical \gls{RG} scale $k$. To be explicit, let us consider a one-parameter family of regulators parameterized by $\xi \in \mathbb{R}_{> 0}$, $r_\xi(y)\coloneqq r(\xi y)$. From the dependence of the flow equation on the regulator, it then follows that the scale-dependent effective actions obtained from calculations with~$r_{\xi=1}$ and~$r_{\xi}$ are directly related:\footnote{This is also correct in cases where only a truncation of the full effective action is considered, such as \gls{LPA} in this subsection.}
\begin{align*}
	\Gamma_k \equiv \Gamma_k^{(\xi=1)} = \Gamma^{(\xi)}_{\sqrt{\xi}k}\,,
\end{align*}
{\it provided} that the initial conditions are chosen accordingly, i.e.,
\begin{align*}
	S\equiv \Gamma_{\Lambda} = \Gamma^{(\xi)}_{\sqrt{\xi}\Lambda}\,.
\end{align*}
Loosely speaking, a change of $\xi$ can be compensated by ``compressing" or ``stretching" the \gls{RG} flow between a given starting point and endpoint. From this, it follows that the results in the \gls{IR} limit~$k\to 0$ do not depend on~$\xi$. In this sense, regulators of the one-parameter regulator family are {\it equivalent}. On the other hand, 
\begin{align*}
	\numberthis
	P^2_{\xi}(y) = \frac{1}{\xi} P^2_{\xi=1}(\xi y) \, ,
\end{align*}
where $P^2_{\xi}(y) = y(1 + r_{\xi}(y))$. If $y_0$ is now a global minimum of $P^2_{\xi}(y)$, then $P^2_{\xi=1}(y)$ has a global minimum at $y=\xi y_0$ and we find $\min_y P^2_{\xi=1}(y) = \xi \min_y P^2_{\xi}(y)$. Thus, any regulator~$R$ of the types considered in this work can be normalized by choosing an appropriate~$\xi$ and hence can be made comparable with respect to \cref{eq:comparability_condition0}. 

With these considerations at hand, we obtain the following order relation from \cref{eq:general_POSS} in \gls{LPA}:
\begin{align*}
	\label{eq:order-relation}
	\mathcal{R}(r_A, r_B) 
	&= \lim_{k \to 0} \frac{l^{(d)}_1(\omega_0(k), r_A)}{l^{(d)}_1(\omega_0(k), r_B)} \numberthis\\
	&= \lim_{\omega \to -1} \frac{l^{(d)}_1(\omega, r_A)}{l^{(d)}_1(\omega, r_B)} \,
\end{align*}
with $\omega_0(k) = \partial^2_\phi U_k (0)/k^2$. Here, we have removed the explicit dependence on~$k$ in the last step, which is possible since the limit $k \to 0$ is equivalent to the limit $\omega \to -1$ according to our choice for~$\omega_{\rm pole}$ above. In the special case of \gls{LPA}, the details of the \gls{RG} flow, i.e., how $\omega_0(k) =  \partial^2_\phi U_k(0)/k^2$ eventually approaches $\omega_{\mathrm{pole}}$, are irrelevant. Thus, no explicit reference \gls{RG} flow and no reference regulator is required for the order relation, in contrast to the general formulation in Eq.~\eqref{eq:general_POSS}. This implies that, in \gls{LPA}, \gls{POSS} effectively only compares the behavior of the threshold functions (for different regulators) close to the singularity with each other.

In \cref{sec:numerical-stability-tests}, we shall indeed demonstrate that a more stable regulator, i.e., a regulator with a stronger self-healing property, generates a more stable \gls{RG} flow which means that the scale-dependent effective potential becomes flat faster in terms of \gls{RG} time in the \gls{IR} limit. More precisely, for $r_A > r_B$, we have $\omega_0(r_A,k) > \omega_0(r_B,k)$ for $k\to 0$, provided that the \gls{RG} flows associated with the two regulators approach the same \gls{IR} limit. Furthermore, we show that an \gls{RG} flow generated by a stable regulator can be solved down to much smaller \gls{RG} scales without numerical breakdowns than it is the case for less stable regulators. In this spirit, more stable regulators generate {\it numerically} more stable \gls{RG} flows. 

Note that the order relation~\eqref{eq:order-relation} depends on the number~$d$ of spacetime dimensions. We shall discuss this dependence in more detail in Subsec.~\ref{sec:construction-of-regulators-preparation}. The independence of the parameter~$m$ follows from L'H$\rm\hat{o}$pital's rule. 

Finally, we would like to add that, for a more general theory including (multiple) bosonic and fermionic degrees of freedom, the order relation effectively boils down to the one which dominates the order relation~\eqref{eq:general_POSS} in the \gls{IR} limit. This is a consequence of the fact that the flow equation for the effective action is simply a sum of the loops associated with the various degrees of freedom.

\section{Construction of regulators}\label{sec:construction-of-regulators}
Following \gls{POSS} and the corresponding order relation, we shall now construct regulators which yield (numerically) stable \gls{RG} flows for studies of simple scalar field theories in \gls{LPA}. 

In the first subsection, we start by discussing general constraints for the regulator shape function~$r$. These constraints can be deduced from the derivation of the Wetterich equation and the properties of the threshold functions discussed in Subsec.~\ref{sec:poss_LPA}. In the second subsection, we aim at providing an intuitive understanding of what actually determines the ``strength'' of the singularity of the threshold function at $\omega_{\mathrm{pole}}=-1$. This is useful since the properties of this singularity control the (numerical) stability of the \gls{RG} flows as we shall demonstrate explicitly in \cref{sec:numerical-stability-tests}. Based on these considerations, we finally construct new regulators in the third subsection and compare them to frequently used ones.

\subsection{Constraints for regulator functions}\label{sec:construction-of-regulators-constraints} 
Let us begin with a discussion of general properties of regulators. The (regulator) function~$R_k(p^2)$ entering the Wetterich equation has to obey the following three conditions~\cite{Wetterich:1992yh}:
\begin{align*}
	\lim_{k\to 0} R_k(p^2) &= 0 \, , \numberthis\label{eq:Rk1}\\
	\lim_{k\to \infty} R_k(p^2) &= \infty \, , \numberthis\label{eq:Rk2} 
\end{align*}
and
\begin{align*}
	\lim_{p\to 0} R_k(p^2) &> 0 \,  \numberthis\label{eq:Rk3}
\end{align*}
for $k>0$. The first condition guarantees that the regulator is removed in the \gls{IR} limit, $k\to 0$, such that we recover the quantum effective action~$\Gamma$. The second condition ensures that the coarse-grained effective action~$\Gamma_k$ approaches the classical action~$S$ in the limit~$k\to \infty$. Finally, the third condition ensures that \gls{IR} divergences in the loop integrals are removed, loosely speaking, by inserting a scale-dependent and momentum-dependent mass term into the propagator. In terms of the regulator shape function~$r$ and the dimensionless variable $y=p^2/k^2$ the three conditions are fulfilled by only two conditions, namely
\begin{align*}
	\numberthis\label{eq:regulator-shape-function-constraints}
	\lim_{y\to 0} y\,r(y) > 0 \, , \quad \lim_{y\to \infty} r(y) = 0 \, .
\end{align*}
Equivalently, in terms of the auxiliary quantity~$P^2$ introduced in \cref{eq:P2}, we have
\begin{align*}
	\numberthis\label{eq:P2-constraints}
	\lim_{y\to 0} P^2(y) > 0 \, , \quad \lim_{y\to \infty} \frac{P^2(y)}{y} = 1 \, .
\end{align*}
We would like to emphasize that the conditions given in \cref{eq:regulator-shape-function-constraints}, or equivalently in \cref{eq:P2-constraints}, are in general not sufficient to guarantee  finiteness of the threshold functions $l^{(d)}_m$ for $\omega > \omega_{\mathrm{pole}}$. Also, these conditions are not sufficient to guarantee that these functions diverge at $\omega = \omega_{\mathrm{pole}}$, see also \cref{sec:construction-of-regulators-preparation}. Therefore, it is necessary to introduce additional constraints for the regulator~$r$ which depend on, e.g., the number $d$ of spacetime dimensions, the parameter $m$, and on the position of the poles of $I^{(d)}_m$. For a more detailed discussion we refer the reader to \cref{app:threshold-function-convergence}, where we also show that the often used exponential regulator $r_{\mathrm{exp}}$ does not yield threshold functions which diverge at $\omega_{\mathrm{pole}}$ in $d > 2$ spacetime dimensions, at least for $m=1$, and hence this regulator spoils the ``self-healing'' property of the \gls{RG} flow discussed in \cref{sec:poss}, in accordance with Ref.~\cite{Pelaez:2015nsa}. From here on, we shall restrict our discussion to regulators~$r$ which yield threshold functions compatible with all the aforementioned constraints.
\begin{figure}
	\includegraphics[width=9cm]{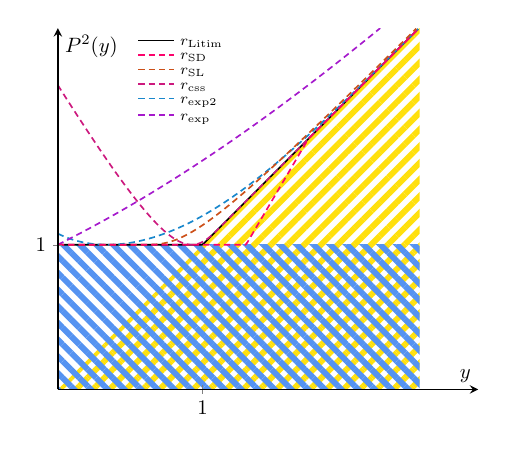}
	\caption{\label{fig:inverse_propagator}%
		Illustration of $P^2(y)$ (corresponding to the regularized squared $d$-momentum in $d$ spacetime dimensions), see \cref{eq:P2}, for a selected set of normalized regulators listed in \cref{tab:regulators}. The black line corresponds to the Litim regulator, see \cref{eq:litim-regulator}. The blue shaded region is forbidden by the normalization condition \eqref{eq:normalization-condition}. In the yellow shaded region the regulator shape function becomes negative, $r(y) < 0$.
	}
\end{figure}

In general, regulators where the threshold function $l^{(d)}_m$ becomes negative for some value of~$\omega$ generate instabilities in the \gls{RG} flow. This can be better understood by looking at the flow equation \eqref{eq:example_LPA_flow} from a fluid-dynamical standpoint~\cite{Grossi:2019urj,Grossi:2021ksl,Koenigstein:2021syz,Koenigstein:2021rxj,Koenigstein:2023wso,Steil:2021cbu}. It then becomes clear that a negative threshold function entails a {\it violation} of the diffusion property of the flow equation, see \cref{eq:diffusion-coefficent} and also the corresponding discussion below. Hence, appropriate regulators should be elements of the set~$\mathcal{A}'$, where 
\begin{align*}
	\numberthis\label{eq:regulator-set-enlarge}
	\mathcal{A}' = \{r \,|\, &\text{$r$ normalized and continuous} \\ &\text{piecewise differentiable}\\&\text{with $l^{(d)}_m(\omega) \geq 0 $ for all $\omega \in (-1, \infty)$}\} \, .
\end{align*}
The integrands of the threshold functions can only become negative if $r'(y) > 0$ for some value of~$y$. Therefore, for simplicity, one may only consider regulators from the set~$\mathcal{A}$ with
\begin{align*}
	\numberthis\label{eq:regulator-set}
	\mathcal{A} = \{r \,|\, &\text{$r$ normalized and continuous} \\ &\text{piecewise differentiable with $r'\leq 0$}\} \subset \mathcal{A}'\, .
\end{align*}
This is the set of regulator functions usually considered in the context of functional \gls{RG} studies.
We add that the restriction to $\mathcal{A}$ also prevents~$P^2$ from entering the yellow shaded region in~\cref{fig:inverse_propagator}, i.e., it prevents~$r$ from becoming negative. Thus, with respect to our normalization condition~\eqref{eq:normalization-condition}, all regulators in $\mathcal{A}$ are ``living" inside the unshaded region in \cref{fig:inverse_propagator} and have at least one point of contact with the line defined by~$P^2(y)=1$. The black line in \cref{fig:inverse_propagator} represents the so-called Litim regulator \cite{Litim:2001up},
\begin{align*}
	\numberthis\label{eq:litim-regulator}
	r_{\mathrm{Litim}}(y)=\left(\frac{1}{y}-1\right) \theta (1-y) \,.
\end{align*}
For this regulator, $P^2$ as a function of~$y$ agrees identically with the boundary of the unshaded region in \cref{fig:inverse_propagator}. Below, we shall show that this regulator is the one which exhibits the strongest singular behavior for regulators in~$\mathcal{A}$.

In the following we shall discuss various examples of regulators which are summarized and categorized in terms of the sets~$\mathcal{A}$ and $\mathcal{A}'$ in \cref{tab:regulators}.
\begin{table}
	
	\begin{ruledtabular}
		\setlength\extrarowheight{8pt}
		\begin{tabular}{l l}
			Regulator & 
			\\
			$r_{\mathrm{exp}}(y)=\frac{1}{\exp(y)-1}$ & Ref.~\cite{Wetterich:1992yh}
			\\
			$r_{\mathrm{mexp}}(y)=\frac{3.92}{e^{\ln(1+3.92) y}-1}$ & Ref.~\cite{Litim:2000ci}
			\\
			$r_{\mathrm{exp2}}(y)=\frac{1}{1-e^{-y-y^2}}-1$ & --
			\\
			$r_{\mathrm{css}}(y)=\frac{c_1}{\exp(c_2 y^b/(1-y^b))-1}\theta\left(1-y\right)$ & Ref.~\cite{Nandori:2012tc}
			\\
			$r_{\mathrm{Litim}}(y)=\left(\frac{1}{y}-1\right) \theta (1-y)$ & Ref.~\cite{Litim:2001up} 
			\\
			$r_{\mathrm{NSL}}(y) = \left(\frac{1}{y}-1\right) \left(\frac{1}{2}\tanh(\epsilon (1-y))+\frac{1}{2}\right)$ &  Eq.~\eqref{eq:naive-smooth-litim-regulator-shape-function}
			\\
			$r_{\mathrm{SL}}(y) = \exp(-\frac{1}{y-\frac{1}{2}}) \theta \left(y-\frac{1}{2}\right) + \frac{1}{y} -1$ & Eq.~\eqref{eq:smooth-litim-regulator-shape-function}
			\\
			$r_{\mathrm{SD}}(y) = \left\{\begin{array}{ll}
				\frac{1}{y} -1 & \text{for $0 \leq y < y_0$}\\[-0em]
				\frac{s(y-y_0) + 1}{y} -1 & \text{for $y_0 \leq y < \frac{1-sy_0}{1-s}$}\\[-0em]
				0 & \text{for $\frac{1-sy_0}{1-s} \leq y$}
			\end{array}\right.$ & Eq.~\eqref{eq:SD-regulator-shape-function}
			\\
			$r_{x_0}(y)=\frac{\left((1-{x_0}) e^{{x_0}-y}+y-1\right) \theta (y-{x_0})+1}{y}-1$ & Eq.~\eqref{eq:hierarchy-of-regulators}
			\\
			$r_{\mathrm{osc}}(y) = \theta(1-y) (A (1-\cos(2 \pi  n y))-y+1)+y$ & Eq.~\eqref{eq:osci_regulator}
		\end{tabular}
	\end{ruledtabular}
	
	\caption{\label{tab:regulators}%
		List of various regulator shape functions. All regulators are in $\mathcal{A}$, except for $r_{\mathrm{SD}}$ and $r_{\mathrm{NSL}}$ which are in~$\mathcal{A}'\setminus\mathcal{A}$. Note that the regulator shape functions $r_{\mathrm{mexp}}$, $r_{\mathrm{exp2}}$, $r_{\mathrm{css}}$, and $r_{\mathrm{NSL}}$ as listed here are not normalized according to Eq.~\eqref{eq:normalization-condition}. 
	}
	
\end{table}

\subsection{Principle of Strongest Singularity\\ and a simple area law}\label{sec:construction-of-regulators-preparation} 
By looking at the definition of the threshold functions in \cref{eq:generic-threshold-function-integrand}, we observe that a larger area under the curve defined by the function~$I^{(d)}_m$ directly corresponds to a stronger singular behavior of the threshold functions close to $\omega_{\mathrm{pole}}=-1$, see \cref{eq:order-relation}. In the following we shall discuss the dependence of the size of this area on $P^2(y)$, i.e., $r(y)$. In particular, we shall show that the minima of $P^2(y)$ play a crucial role since they essentially determine the size of the area in the limit $\omega \to -1$ and therefore the strength of the singular behavior of the associated threshold function.

Let us begin by considering a normalized regulator which is at least continuously differentiable \emph{once} and has a unique global minimum of $P^2(y)$ at $y_0 \in (0, 1)$. At $y_0$, the value of the regulator as well as its derivative agree with the corresponding ones of the Litim regulator by construction:

\begin{align*}
	\numberthis\label{eq:introofy0}
	r(y_0) = r_{\mathrm{Litim}}(y_0) \quad \text{and} \quad r'(y_0) = r'_{\mathrm{Litim}}(y_0) \, .
\end{align*}
Thus, the integrand of the generic threshold function defined in~\cref{eq:generic-threshold-function-integrand} agrees with the one obtained from the Litim regulator at $y_0$ (for any number of spacetime dimensions), i.e., we have
\begin{align*}
	\numberthis\label{eq:intergand-contact-point}
	&I^{(d)}_m(\omega, r(y_0), r'(y_0), y_0) \\
	&\quad = I^{(d)}_m(\omega, r_{\mathrm{Litim}}(y_0), r'_{\mathrm{Litim}}(y_0), y_0) \, .
\end{align*}
Close to $y_0$, i.e., for sufficiently small $\epsilon > 0$, we find
\begin{align*}
	\numberthis\label{eq:estimate1}
	r(y_0 \pm \epsilon) &> r_{\mathrm{Litim}}(y_0 \pm \epsilon) \, , \\
	\numberthis\label{eq:estimate2}
	r'(y_0 + \epsilon) &> r'_{\mathrm{Litim}}(y_0 + \epsilon) \, , \\
	\numberthis\label{eq:estimate3}
	r'(y_0 - \epsilon) &< r'_{\mathrm{Litim}}(y_0 - \epsilon) \, .
\end{align*}
This immediately implies that
\begin{align*}
	\numberthis
	&I^{(d)}_m(\omega, r(y_0 + \epsilon), r'(y_0 + \epsilon), y_0 + \epsilon) \\
	&\quad< I^{(d)}_m(\omega, r_{\mathrm{Litim}}(y_0 + \epsilon), r'_{\mathrm{Litim}}(y_0 + \epsilon), y_0 + \epsilon) \, .
\end{align*}
Since we assumed that $r(y)$ has only one point of contact with the Litim regulator shape function at $y_0 \in (0, 1)$, we can conclude that
\begin{align*}
	\numberthis
	&I^{(d)}_m(\omega, r(y), r'(y), y) < I^{(d)}_m(\omega, r_{\mathrm{Litim}}(y), r'_{\mathrm{Litim}}(y), y)
\end{align*}
for all $y \in (y_0, 1]$. For values $y \in [0, y_0)$, we can in general not find a corresponding inequality. There, the denominator and the numerator of the integrand~\eqref{eq:generic-threshold-function-integrand} nontrivially compete with each other, such that the integrand can exceed the values obtained with the Litim regulator. 
\begin{figure*}
	\includegraphics[]{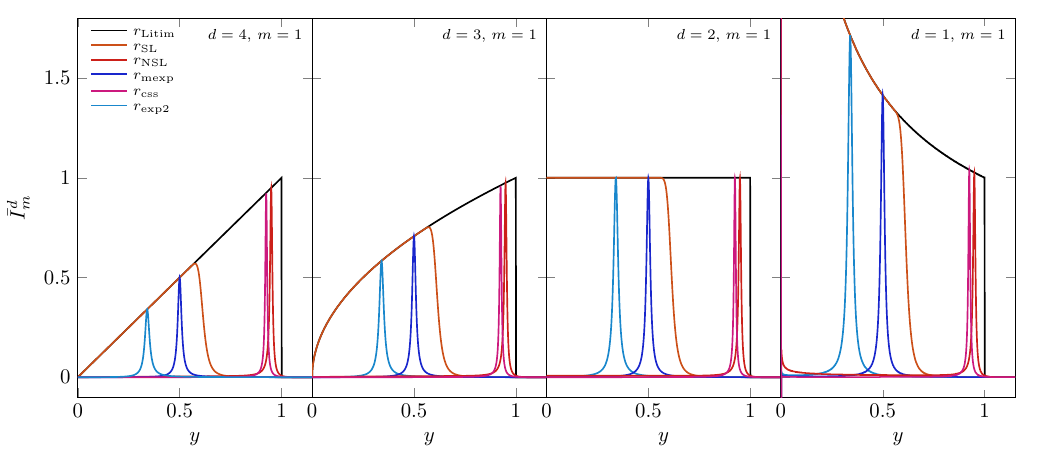}
	\caption{\label{fig:threshold_function_integrand}%
		The auxiliary function $\bar I^{(d)}_m$ for $m=1$ at $\omega=-0.9999$ for different numbers of spacetime dimensions~$d$. For a definition of the various regulators, see \cref{tab:regulators}.}
\end{figure*}
However, in the limit $\omega \to -1$, the latter is not the case anymore. To show this, it is convenient to introduce the following auxiliary function:
\begin{align*}
	\numberthis\label{eq:lbardef}
	\bar I^{(d)}_m \coloneqq (1+\omega)^m I^{(d)}_m\,.
\end{align*}
This function is no longer singular and therefore the integration with respect to $y$ yields a function which is well-defined in the limit $\omega \to -1$. For any $y$ with $P^2(y) > 1$, $\bar I^{(d)}_m$ is suppressed and even approaches zero for~$\omega\to -1$. Only points with $P^2(y) = 1$ (e.g., at $y=y_0$) have a finite non-zero limit, which is
\begin{align*}
	\numberthis
	-y_0^{d/2+1}r'_{\mathrm{Litim}}(y_0) &= y_0^{d/2-1} \,.
\end{align*}
This behavior is illustrated in \cref{fig:threshold_function_integrand}, where we show $\bar I^{(d)}_1$ for $\omega = -0.9999$ for various regulators. In the limit $\omega \to -1$, the area under the curve defined by the integrand associated with the threshold function obtained with the Litim regulator indeed exceeds the one of any other regulator shown in \cref{fig:threshold_function_integrand}.

In \cref{fig:threshold_function_integrand}, we also observe the dependence of the order relation $\mathcal{R}$ on the number of spacetime dimensions. In order to maximize the area under the curve defined by the integrand of the threshold function, $y_0$ should be close to $y=1$ for $d>2$ and close to $y=0$ for $d<2$, see~\cref{fig:threshold_function_integrand}. Furthermore, if the regulator ``lies" closer to the line $P^2(y) = 1$ at~$y_0$, e.g., $\partial_y^2 {P^2}(y)|_{y=y_0} \approx 0$, the area under the curve defined by the integrand increases.
Comparing the areas associated with various regulators and thus the strength of the singular behavior for~$\omega\to -1$, we deduce from \cref{fig:threshold_function_integrand} for $d = 4$ that
\begin{align*}
	\numberthis\label{eq:hierarea}
	&r_{\mathrm{Litim}}(y)>r_{\mathrm{SL}}(y)>r_{\mathrm{NSL}}(y) \nonumber\\
	&\qquad\qquad\qquad >r_{\mathrm{mexp}}(y)>r_{\mathrm{css}}(y)>r_{\mathrm{exp2}}(y) \, .
\end{align*}

In \cref{fig:threshold_function_integrand}, we also observe that, with decreasing number of spacetime dimensions, the area under the curve defined by the integrand $\bar I^{(d)}_m$ increases.\footnote{Loosely speaking, this behavior of the integrand tends to counteract the formation of a nontrivial ground state of the potential, as associated with spontaneous symmetry breaking. However, as we shall discuss in \cref{sec:numerical-stability-tests}, this is not the reason for the absence of spontaneous symmetry breaking in low dimensions, i.e., $d < 2$, as the latter is a regularization-independent statement~\cite{PhysRevLett.17.1133,PhysRev.158.383,ColemanSSB}.}

In general, we conclude that, for normalized regulators, every point $y_0$ with 
\begin{align*}
	\numberthis\label{eq:def-C}
	y_0 \in \mathcal{C} \coloneqq \{ y \in [0, \infty) \,|\, P^2(y) = 1\}
\end{align*}
generates a peak in the auxiliary function~$\bar I^{(d)}_m$ with the height
\begin{align*}
	\numberthis
	-y_0^{d/2+1}r'(y_0)
\end{align*}
for $\omega \to -1$. In the following we shall refer to elements in $\mathcal{C}$ as {\it contact points}. For regulators with a finite set $\mathcal{C}$, the auxiliary function
\begin{align*}
	\numberthis
	\bar l^{(d)}_m(\omega) = \int_0^{\infty} \d y\, \bar I^{(d)}_m(\omega, r(y), r'(y), y)\,,
	\label{eq:AuxTF}
\end{align*}
which is directly related to the threshold function~$l^{(d)}_m(\omega)$ via \cref{eq:lbardef}, tends to zero for $\omega \to -1$.\footnote{Note that even if $\bar l^{(d)}_m$ vanishes for $\omega \to -1$, $l^{(d)}_m$ can still diverge in this limit.} To obtain a non-zero, finite limit, it is therefore necessary to use a regulator with an infinite set $\mathcal{C}$.
A maximization of the area under the curve $\bar I^{(d)}_m$ and thus an increase of the strength of the singular behavior of the threshold function~$l^{(d)}_m(\omega)$ close to $\omega = -1$ can be achieved by increasing the set $\mathcal{C}$.
Apparently, according to our definition, {\it stable} regulators should be those which have a large set $\mathcal{C}$ or even an {\it infinite} set $\mathcal{C}$ such as the Litim regulator.

We would like to emphasize that, with the order relation~\eqref{eq:order-relation} at hand, it directly follows that, in \gls{LPA}, the Litim regulator is the {\it unique optimal} regulator in $\mathcal{A}$ with respect to \gls{POSS} since $\mathcal{C} = [0, 1]$ and $r'\leq 0$. Thus, in \gls{LPA}, \gls{POSS} uniquely singles out the same optimal regulator as the functional optimization discussed in Ref.~\cite{Pawlowski:2005xe}. If we consider regulators in $\mathcal{A}'$, an optimal regulator $r_{\mathrm{opt}}$ with respect to \gls{POSS} has to fulfill:
\begin{align*}
	\numberthis\label{eq:optimal_regulator_in_A_prime}
	\bar l^{(d)}_1(\omega = -1,r_{\mathrm{opt}}) = \max_{r \in \mathcal{A}'}\,\bar l^{(d)}_1(\omega = -1,r) \, .
\end{align*}
\subsection{Examples of regulators}\label{sec:construction-of-regulators-construction}
The \acrlong{POSS} is not only useful to order and compare existing regulators, it also provides us with an intuitive prescription to construct {\it new} regulators which allow to obtain (numerically) stable \gls{RG} flows. We now aim at such a construction for studies in \gls{LPA}. In the first part of this subsection we shall focus on the construction of regulators in the set~$\mathcal{A}$ which are {\it differentiable} but still come with a singularity strength comparable to the one of the Litim regulator.
The second part then deals with regulators in the set~${\mathcal A}^{\prime}$. The singular behavior of the regulators in this set can even be stronger than the one of the Litim regulator as we shall show by an explicit construction of such a regulator.

As already discussed above, the Litim regulator is the unique optimal regulator in $\mathcal{A}$ with respect to \gls{POSS}. Therefore, it represents the natural starting point for the construction of new (at least ``almost optimal") differentiable regulators. For example, if we would like to have a differentiable Litim-like regulator, we may be tempted to replace the Heaviside function in the definition of the Litim regulator \eqref{eq:litim-regulator} by a ``smeared-out" version of it, e.g., $\theta(x) \approx \tfrac{1}{2}(1 + \tanh(\epsilon x))$ with some large value of $\epsilon$, e.g., $\epsilon = 10$, yielding the \gls{NSL} regulator,
\begin{align*}
	\numberthis\label{eq:naive-smooth-litim-regulator-shape-function}
	r_{\mathrm{NSL}}(y) = \left(\frac{1}{y}-1\right) \left(\frac{1}{2}\tanh(\epsilon (1-y))+\frac{1}{2}\right)\,.
\end{align*}
However, after a normalization of this regulator according to \cref{eq:normalization-condition}, it then turns out that it has only one contact point, \mbox{$\mathcal{C} = \{y_0\}$}. This implies that the resulting threshold function only exhibits a weak singular behavior at~$\omega=-1$ as illustrated in \cref{fig:threshold_function_integrand}. Hence, this simple ``smeared-out version" of the Litim regulator $r_{\mathrm{NSL}}(y)$ is far from being a good choice with respect to \gls{POSS}. Note also that $r_{\mathrm{NSL}}$ is not even contained in $\mathcal{A}$ since it enters the yellow shaded region in \cref{fig:inverse_propagator}. 

A \gls{SL} regulator with an infinite set~$\mathcal{C}$ is given by, e.g.,
\begin{align*}
	\numberthis\label{eq:smooth-litim-regulator-shape-function}
	r_{\mathrm{SL}}(y) = \exp(-\frac{1}{y-\frac{1}{2}}) \theta \left(y-\frac{1}{2}\right) + \frac{1}{y} -1 \, ,
\end{align*}
which yields
\begin{align*}
	\numberthis
	P^2_{\mathrm{SL}}(y) = y \exp \left(-\frac{1}{y-\frac{1}{2}}\right) \theta \left(y-\frac{1}{2}\right)+1 \, ,
\end{align*}
see \cref{fig:threshold_function_integrand} for an illustration.\footnote{Before employing the regulator~\eqref{eq:smooth-litim-regulator-shape-function} in a specific study, it is required to ensure that this regulator also fulfills the corresponding requirements detailed in \cref{app:threshold-function-convergence}.}

At first glance, the regulator~\eqref{eq:smooth-litim-regulator-shape-function} appears to be very similar to the \gls{CSS} regulator shape functions introduced in Ref.~\cite{Nandori:2012tc},  see \cref{tab:regulators} for the definition of the corresponding shape function~$r_{\mathrm{css}}(y)$. For concrete applications of this class of regulators, we refer the reader to Refs.~\cite{Marian:2013zza, Nandori:2013nda, Kovacs:2014mia}. However, these regulators are constructed such that $r(y) \neq 0$ for a compactly supported region which is not the case for the regulator in \cref{eq:smooth-litim-regulator-shape-function}. In fact, the regulators constructed in Ref.~\cite{Nandori:2012tc} have only one contact point, $\mathcal{C}=\{y_0\}$ (see \cref{fig:inverse_propagator}), and are therefore unstable in the spirit of \gls{POSS}. 

Given the fact that the regulator \eqref{eq:smooth-litim-regulator-shape-function} comes with an infinite set~$\mathcal{C}$ (see \cref{fig:inverse_propagator}), this regulator represents a reasonable starting point for the construction of other new regulators which are infinitely differentiable on the full domain and even close to the Litim regulator with respect to \gls{POSS}. Such regulators may potentially then also turn out to be useful for studies beyond \gls{LPA} where differentiability of regulators in general becomes a crucial property at some point. Furthermore, it is in principle also possible to construct regulators which combine the properties of the ones associated with the shape functions $r_{\mathrm{SL}}(y)$ and $r_{\mathrm{css}}(y)$, i.e., compactly supported smooth regulator shape functions which approach the Litim regulator at a nonzero~$y$.\footnote{More precisely, we may construct a smooth regulator with $P^2(y) = 1$ for $0 \leq y < y_0$ and $P^2(y)=y$ for $y_1 \leq y$. The intermediate region, $y_0 \leq y < y_1$ can then be bridged by a smooth function which smoothly connects to the functions in the other regions. It may be chosen to be of, e.g., the following form:
\begin{eqnarray}
	P^2(y) &=& y \exp{-a\,\exp{\frac{1}{y-y_0} 	+\frac{1}{y-y_1}}} \nonumber \\
    && \qquad -\exp{-\exp{\frac{1}{y-y_0}+\frac{1}{y-y_1}}}+1 \,, \nonumber
\end{eqnarray}
where $y_0 = 0.5$, $y_1 = 1.5$ and $a = 0.625$.} Such a combination may be useful for studies beyond \gls{LPA}, i.e., at high(er) orders of the derivative expansion of the effective action, where, e.g., the Litim regulator is no longer optimal in the spirit of functional optimization~\cite{Pawlowski:2005xe} and, depending on the order of the derivative expansion, not even applicable because of the lack of differentiability.
 
As a second application of \gls{POSS}, we construct a regulator whose singular behavior even exceeds the one of the Litim regulator. Clearly, this is not possible within the set~$\mathcal{A}$ since, in \gls{LPA}, the Litim regulator is the unique optimal regulator in~$\mathcal{A}$ with respect to \gls{POSS}. Therefore, we have to consider the set~$\mathcal{A}' \setminus \mathcal{A}$.\footnote{It would be interesting to include regulators from the set ${\mathcal A}^{\prime} \setminus {\mathcal A}$ into an analysis based on functional optimization as discussed in Ref.~\cite{Pawlowski:2005xe}.} A simple initial guess for such a regulator with a \gls{SD} behavior may be  
\begin{align*}
	\numberthis\label{eq:SD-regulator-shape-function}
	r_{\mathrm{SD}}(y) = \begin{cases}
		\frac{1}{y} -1\ &\quad (0 \leq y < y_0)\\
		\frac{s(y-y_0) + 1}{y} -1  &\quad (y_0 \leq y < \frac{1-sy_0}{1-s})\\
		0 &\quad (\frac{1-sy_0}{1-s} \leq y)
	\end{cases} \, ,
\end{align*}
which yields
\begin{align*}
	P^2_{\mathrm{SD}}(y) = \begin{cases}
		1  &\quad (0 \leq y < y_0)\\
		s(y-y_0) + 1  &\quad (y_0 \leq y < \frac{1-sy_0}{1-s})\\
		y  &\quad (\frac{1-sy_0}{1-s} \leq y)
	\end{cases} \,.
\end{align*}
Here, $y_0 > 1$ denotes the width of the interval $\mathcal{C} = [0, y_0]$ and $s > 1$ is the slope of the linear connection line between $P^2(y=y_0) = 1$ and the point $y=(1-sy_0)/(1-s)$ above which we have $P^2(y)=y$. The parameters~$s$ and~$y_0$ are constrained by the fact that we are requiring the threshold functions to be positive. Note that the regulator assumes negative values between $y=1$ and $y=(1-sy_0)/(1-s)$, i.e., it enters the yellow shaded region in \cref{fig:inverse_propagator}. In $d=4$, we find an upper bound for $y_0$, which is $y_0 < 3/2$. For any such~$y_0$, a lower bound for $s$ exists such that the threshold functions in \gls{LPA} remain positive. For example, in $d=4$, we may choose $y_0= 1.3$ and $s=1.7$. Note that the simple structure of this regulator in principle allows for an analytic computation of the threshold functions. 

\section{Stability of \gls{RG} flows}\label{sec:numerical-stability-tests}
In order to illustrate that \gls{POSS} is indeed useful for the construction of regulators which generate (numerically) stable \gls{RG} flows, we consider again the \gls{RG} flow equation~\eqref{eq:example_LPA_flow} for the effective potential $U_k$ of a scalar field theory in $d=4$ spacetime dimensions in \gls{LPA} for various regulators. 
To be specific, we consider several regulators including the standard Litim regulator ($r_{\mathrm{Litim}}$), the naively ``smeared-out" Litim regulator ($r_{\mathrm{NSL}}$), a compactly supported smooth regulator ($r_{\mathrm{css}}$), two variants of exponential regulators ($r_{\mathrm{exp2}}$ and $r_{\mathrm{mexp}}$), the smooth regulator ($r_{\mathrm{SL}}$) introduced in \cref{eq:smooth-litim-regulator-shape-function}, and the \gls{SD} regulator ($r_{\mathrm{SD}}$) introduced in \cref{eq:SD-regulator-shape-function} associated with a particularly strong singular behavior close to~$\omega=-1$, see also \cref{tab:regulators} for the definitions of these regulators. Moreover, in order to demonstrate explicitly the relation between the stability of the \gls{RG} flows and the set~$\mathcal C$ of a given regulator,\footnote{Recall that the set~$\mathcal C$ contains all points of contact of a given regulator with $P^2(y) = 1$, see \cref{eq:def-C}.} we add yet another family of regulators:
\begin{align*}
	\numberthis\label{eq:hierarchy-of-regulators}
	r_{x_0}(y) = \frac{\left((1-{x_0}) e^{{x_0}-y}+y-1\right) \theta (y-{x_0})+1}{y}-1 \,.
\end{align*}
This regulator family can be adjusted by choosing \mbox{$x_0 \in [0, 1]$}. For $y \in [0, x_0]$, we have $P^2(y) = 1$ and, for $y \in [x_0, \infty]$, we find that $P^2(y)$ approaches the asymptote~$y$ exponentially. For $x_0=1$, we find $r_{x_0}(y) = r_{\mathrm{Litim}}(y)$. Thus, by adjusting $x_0$, we can continuously vary the set~$\mathcal{C}$ which implies a variation of the strength of the singular behavior of the threshold function. Moreover, to illustrate the effect of having more than one but still a finite number of global minima of $P^2(y)$, we introduce an oscillatory regulator ($r_{\mathrm{osc}}$): 
\begin{align*}
	\numberthis\label{eq:osci_regulator}
	r_{\mathrm{osc}}(y) = \theta(1-y) (A (1-\cos(2 \pi  n y))-y+1)+y \, ,
\end{align*}
where $n = |\mathcal{C}| - 1$ and $A < 1/(2 \pi  n (1-\frac{3}{4 n})-1)$ such that $r_{\mathrm{osc}}'(y) \leq 0$. For concreteness, we shall use $A \approx 0.0157$ and $n = 10$ in our numerical studies below. Note that~$r_{\mathrm{osc}}$ as well as the regulator family~$r_{x_0}$ do not fall in the class of differentiable regulators.

\begin{figure}
	\includegraphics[]{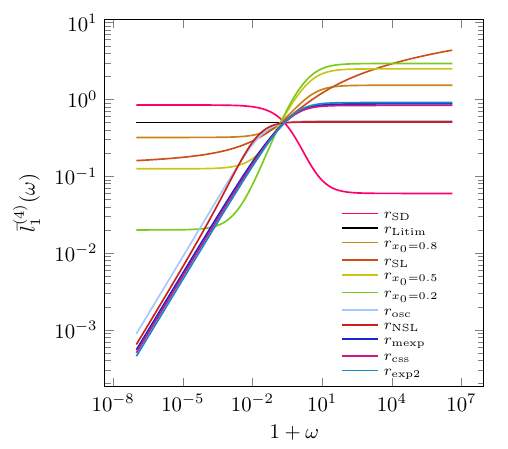}
	\caption{\label{fig:regulator_hierarchy} 
		 Dependence of the auxiliary function $\bar l^{(4)}_{1}$ on the mass parameter~$\omega$. The definition of this function, which is directly related to the threshold function~$l^{(4)}_{1}$, can be found in \cref{eq:AuxTF}. Note that any regulator with an infinite set $\mathcal{C}$ approaches a nonzero value for $\omega \to -1$. From the $\omega$-dependence of these functions and the order relation~\eqref{eq:order-relation}, we can extract the hierarchy of regulators given in \cref{eq:regulator-hierarchy}.
	}
\end{figure}

For the regulators considered in this section, \gls{POSS} predicts the following regulator hierarchy for \gls{LPA} in~\mbox{$d=4$} spacetime dimensions:
\begin{align*}
	&r_{\mathrm{SD}}(y)>r_{\mathrm{Litim}}(y)>r_{x_0=0.8}(y)>r_{\mathrm{SL}}(y)\\
	&\quad>r_{x_0=0.5}(y)>r_{x_0=0.2}(y)>r_{\mathrm{osc}}(y) \\
	&\quad>r_{\mathrm{NSL}}(y)>r_{\mathrm{mexp}}(y)>r_{\mathrm{css}}(y)>r_{\mathrm{exp2}}(y) \,.
	\numberthis\label{eq:regulator-hierarchy}
\end{align*}
This follows from \cref{fig:regulator_hierarchy} together with \cref{eq:order-relation}. More precisely, for regulators with an infinite set $\mathcal{C}$, the auxiliary function $\bar l^{(4)}_1$ approaches a finite value in the limit~$\omega\to -1$, whereas it approaches zero for those regulators with a finite set $\mathcal{C}$, see \cref{fig:regulator_hierarchy}. Note that, in accordance with \gls{POSS}, we find that the oscillatory regulator $r_{\mathrm{osc}}$ with more than one element in $\mathcal{C}$ is {\it more stable} than any of the analytic regulators with only one contact point, i.e., $\mathcal{C} = \{y_0\}$. Moreover, from the standpoint of \gls{POSS}, the \gls{SD} regulator is {\it more stable} than the Litim regulator.
However, we emphasize that these two regulators are associated with two different sets of regulators, namely the sets~${\mathcal A}^{\prime}\setminus\mathcal{A}$ and~${\mathcal A}$, respectively.

In order to solve the flow equation for the effective potential numerically, we follow Refs.~\cite{Aoki:2014,Aoki:2017rjl,Grossi:2019urj,Grossi:2021ksl,Koenigstein:2021syz,Koenigstein:2023wso} and reformulate the \gls{RG} flow equation \eqref{eq:example_LPA_flow} such that it resembles the form of conservation laws in fluid dynamics. To be specific, we simply take a total derivative on both sides with respect to $\phi$ and introduce the slope $u_k(\phi) = \partial_\phi U_k(\phi)$ of the effective potential with respect to the field~$\phi$ as a new variable. The flow equation for the latter quantity then reads
\begin{align*}
	\numberthis\label{eq:flow-equation}
	\partial_t u_k(\phi) &=  \frac{\d}{\d\phi} \left( -\frac{1}{2}\frac{\mathrm{surf}(d) k^d}{(2\pi)^d}\, l^{(d)}_{1}(\omega) \right) \nonumber \\
	&= \frac{1}{2}\frac{\mathrm{surf}(d) k^{d-2}}{(2\pi)^d}\, l^{(d)}_{2}(\omega) \,  \partial^2_\phi u_k(\phi) \,.
\end{align*}
This flow equation can be viewed as a non-linear diffusion equation, where
\begin{align*}
	\numberthis\label{eq:diffusion-coefficent}
	\alpha_k(\omega) = \frac{1}{2}\frac{\mathrm{surf}(d) k^{d-2}}{(2\pi)^d}\, l^{(d)}_{2}(\omega) \geq 0
\end{align*}
is the diffusion coefficient.

The flow equation~\eqref{eq:flow-equation} can now be solved with methods borrowed from numerical fluid dynamics~\cite{Koenigstein:2021syz,Koenigstein:2023wso} without relying on a specific ansatz for the effective potential.
As numerical method for the solution of the \gls{RG} flow equation~\eqref{eq:flow-equation}, we employ the \gls{KT} scheme~\cite{KTO2-0}. This is a finite-volume method which has already been tested and used to solve such equations, see, e.g., Ref.~\cite{Koenigstein:2021rxj}.\footnote{The precise implementation of the \gls{KT} scheme in the context of functional \gls{RG} equations used in the present work can be found in Refs.~\cite{Koenigstein:2021syz,Stoll:2021ori}. For a meaningful comparison of the \gls{RG} flows with different regulators, we employed the same numerical control parameters: $\phi_{\mathrm{max}} = 1.5 \Lambda$, $\Delta\phi = 0.001 \Lambda$, for all regulators. As numerical time stepper we used {\it solve\_ivp} \cite{2020SciPy-NMeth} with ``LSODA'' with $rtol = atol = 10^{-10}$ for its relative and absolute error.} Note that a negative diffusion coefficient~\eqref{eq:diffusion-coefficent} would lead to backwards diffusion which causes numerical instabilities at least within our numerical treatment. 

Since we are interested in studying situations where the ground state of the theory is governed by spontaneous symmetry breaking in the \gls{IR} limit, we briefly discuss the associated emergence of a nontrivial minimum in the effective potential by examining the flow equation~\eqref{eq:flow-equation}. From this equation, we deduce that spontaneous symmetry breaking can only appear if the slope $u_k(\phi)$ of the effective potential~$U_k(\phi)$ with respect to~$\phi$ splits up into two distinct regions, separated by an ``insulating point" $\phi_0$ with $\phi_0^2 > 0$. This insulating point is defined by $\omega = \partial_{\phi}u_k(\phi=\phi_0)/k^2 = 0$.\footnote{Note that~$\phi_0$ should not be confused with the minimum of the effective potential.} At this point, the diffusion coefficient $\alpha_k(0)$ vanishes in the \gls{IR} limit. For example, in $d > 2$, the diffusion coefficient vanishes for $k\to 0$, since $l^{(2)}_{2}(0)$ is a constant and thus
\begin{align*}
	\numberthis
	\alpha_k(0) \propto  k^{(d-2)}\,.
\end{align*}
Hence, for $d > 2$, it is at least in principle possible to encounter spontaneous symmetry breaking in the \gls{IR} limit. On the other hand, for~$d<2$, $\phi_0$ is always affected by a non-vanishing diffusion coefficient such that it cannot be an insulating point and therefore the dynamics of bosons as described by our scalar field tends to restore the symmetry in the \gls{IR} limit.

Let us now assume that a finite insulating point~$\phi_0$ exists in the \gls{IR} limit. In the region $\phi^2 < \phi_0^2$, where the slope of $u_k(\phi)$ is negative (i.e., $\omega < 0$), the diffusion coefficient becomes much greater than in the region $\phi^2 > \phi_0^2$ (i.e., $\omega > 0$) at sufficiently small \gls{RG} scales, since every point inside the region $\phi^2 < \phi_0^2$ is ``pushed" into the singularity associated with \mbox{$\omega = -1$}. As a consequence, this region becomes flat and does not contain any physical information in the limit~$k\to 0$.

For the \gls{RG} flow equation \eqref{eq:flow-equation} with $d=4$, a finite insulating point potentially exists for certain classes of initial conditions at the scale~$k=\Lambda$, e.g.,
\begin{align*}
	\numberthis\label{eq:uvpotential}
	U_{\Lambda}(\phi) = \frac{1}{2}m^2 \phi^2 + \frac{1}{4} \lambda \phi^4\,
\end{align*}
with~$m^2<0$ and~$\lambda >0$. In the following, we set \mbox{$\lambda = 1.0$} for all regulators. The parameter $m^2 < 0$ is tuned such that we obtain $\phi_{\mathrm{min}}/ \Lambda = 0.3$ for the position of the minimum of the effective potential at the scale at which the \gls{RG} flow breaks down numerically for a given regulator.\footnote{Note that the (initial) values for $m^2/\Lambda^2$ depend only mildly on the regulators considered in the present work. To be more specific, we find $m^2/\Lambda^2 \in [-0.099,-0.093]$.} In this way, we ensure that our comparison of the stability of \gls{RG} flows resulting from different regulators is at least to some extent meaningful. Of course, ideally, such a comparison would require to tune the initial conditions such that the effective potential~$U_k$ is identical for~$k\to 0$ for all regulators. However, this is impossible as all \gls{RG} flows eventually break down at a finite \gls{RG} scale in {\it our} present numerical setup due to limited precision of the time stepper.\footnote{Note that, as discussed in Ref.~\cite{Ihssen:2023qaq}, the \gls{RG} flow equation for the effective potential can be reformulated to circumvent the issue of a numerical breakdown at a finite \gls{RG} scale. At least for a certain class of regulators, this suggests that it is possible to find a reformulation of the flow equation or an improvement of the numerical solver which potentially allows to follow the \gls{RG} flow down to very small \gls{RG} scales where the potential eventually becomes flat. However, our present work focuses on the construction of regulators leading to a fast flattening of the effective potential in terms of \gls{RG} time which we demonstrate with a fixed numerical setup. Such a fast flattening of the effective potential essentially introduces a natural stabilization of \gls{RG} flows as discussed above. In any case, a combination of the reformulation presented in Ref.~\cite{Ihssen:2023qaq} with regulators considered optimal according to our definition would be interesting but is beyond the scope of the present study.}

In order to illustrate that the hierarchy of regulators~\eqref{eq:regulator-hierarchy} indeed provides us with a {\it good} measure for the stability of \gls{RG} flows, we solve the flow equation~\eqref{eq:flow-equation} for the different regulators and extract $\omega_{\mathrm{min}}(k)$, which is defined to be the smallest value of $\omega$ at a given \gls{RG} scale~$k$. This quantity then allows us to illustrate how fast (in terms of the \gls{RG} time~$t$) the \gls{RG} flow approaches the singularity.
\begin{figure}
	\includegraphics[]{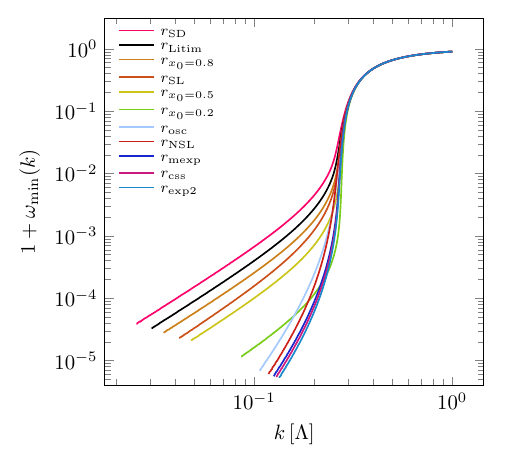}
	\caption{\label{fig:stability_test}%
		The quantity $\omega_{\mathrm{min}}(k)$ as obtained from various regulators as a function of the \gls{RG} scale~$k$, down to the scale where the numerical calculations break down for a given regulator. The ordering of these breakdown scales correspond to the regulator hierarchy predicted by \gls{POSS}, see \cref{eq:regulator-hierarchy} and also \cref{fig:regulator_hierarchy}. Note that, at a fixed \gls{RG} scale $k$, a larger $\omega_{\mathrm{min}}$ corresponds to a flatter potential. This implies that, in terms of \gls{RG} time, the effective potential becomes flat faster for regulators which are considered to be preferred according to \gls{POSS}. The derivative of the effective potential, $u_k(\phi) = \partial_\phi U_k(\phi)$, at the different (numerical) breakdown scales is shown in \cref{fig:u_plot}.
	}
\end{figure}

\begin{figure}
	\includegraphics[]{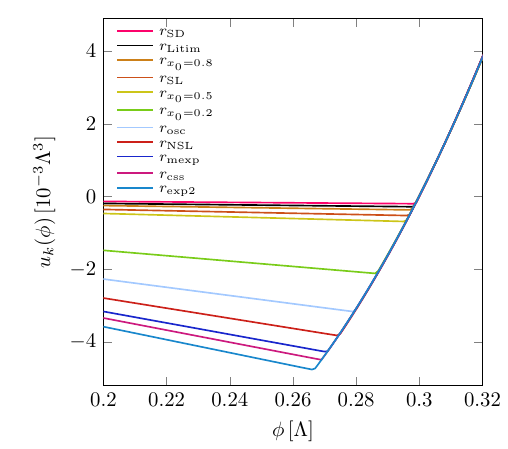}
	\caption{\label{fig:u_plot}%
		Derivative of the effective potential, $u_k(\phi)=\partial_\phi U_k(\phi)$, as a function of~$\phi$ for various regulators at the scale~$k_{\text{break}}$, where the \gls{RG} flow breaks down. The \gls{UV} potentials of the \gls{RG} flows associated with the various regulators are tuned such that the position~$\phi_{\text{min}}$ of the minimum of the effective potential is the same for all regulators at the respective scale~$k_{\text{break}}$, i.e.,~$\phi_{\text{min}}/\Lambda=0.3$. Note that the position of the minimum of the potential is associated with the position of the zero of the depicted derivative of the effective potential, $u_k(\phi)$. We observe that the hierarchy of regulators predicted by \gls{POSS} is in accordance with the observed ``flatness" of the effective potentials indicated by a small/vanishing derivative $u_k(\phi)$, see also \cref{eq:regulator-hierarchy}.}
\end{figure}

For the regulators considered in this work, $\omega_{\mathrm{min}}$ is shown in \cref{fig:stability_test}. We observe that, at a fixed, sufficiently small \gls{RG} scale, the values of $\omega_{\mathrm{min}}$ follow identically the regulator hierarchy predicted by \gls{POSS}, see \cref{eq:regulator-hierarchy}: More stable \gls{RG} flows yield flatter potentials, which is a direct consequence of the self-healing property. Consequently,
less stable \gls{RG} flows approach the singularity at $\omega=\omega_{\mathrm{pole}} = -1$ faster in terms of the \gls{RG} time~$t$ and eventually break down earlier due to the limited numerical precision, in accordance with our definition of numerical stability. Also, the ordering of the break down scales associated with the various \gls{RG} flows is in perfect agreement with the predicted regulator hierarchy as given in \cref{eq:regulator-hierarchy}. As it should be, this ordering is also visible in \cref{fig:u_plot}, where we show the slope $u_{k}(\phi)$ of the effective potential at the numerical breakdown scale for the regulators considered in our analysis.\footnote{The numerical stability as well as the potentials at the breakdown scale presented in \cref{fig:u_plot} do not rely on the normalization condition~\eqref{eq:normalization-condition}, since the (abstract) \gls{RG} trajectory is not affected by a rescaling and hence the numerical solver will break down at the same infrared potential.}

Comparing the \gls{RG} flows obtained from the various regulators, we observe that the values of the smallest \gls{RG} scale, which is numerically accessible for a given regulator, differ significantly. In fact, the breakdown scales for regulators with only one element in~${\mathcal C}$ are a factor of $5-10$ times greater than the scales that can be reached with regulators with an infinite set~${\mathcal C}$.

In summary, our analysis provides a quantitative confirmation that the (numerical) stability of \gls{RG} flows can indeed be improved by a suitable choice of the regulator. \gls{POSS} allows to single out regulators which generate (numerically) stable \gls{RG} flows. Such regulators allow to access the deep infrared regime, where the potential becomes convex. From a  phenomenological standpoint, the access to this regime is ultimately required for quantitative studies of critical behavior in the presence of a (small) explicit symmetry breaking~\cite{Braun:2023qak} and the computation of the corresponding scaling functions, as done in, e.g., Ref.~\cite{Braun:2010vd}. For such studies, the use of regulators with an infinite set of contact points appears inevitable.

\section{Conclusions}\label{sec:conc}
In this work, we discussed the optimization of functional \gls{RG} flows from the standpoint of stability which is tightly connected to the presence of zeroes in the spectrum of the regularized two-point function. In fact, the inverse of the regularized two-point function determines the \gls{RG} flow of the effective action and therefore the zero eigenvalues of the regularized two-point function define singular points which control how fast (measured in terms of \gls{RG} time) the effective potential becomes convex, in particular in situations where the dynamics of a theory is governed by spontaneous symmetry breaking.

We discussed that the stability in numerical calculations can indeed be considerably improved by a suitable choice of the regulator. Of course, such a stability analysis requires to compare different \gls{RG} flows which, within a given truncation, corresponds to comparing different regulators. To render such a comparison meaningful, we introduced a comparability condition and a corresponding principle, the \acrlong{POSS}, which provides us with an order relation for regulator functions.

We demonstrated the application of this general principle by using the example of a scalar field theory in \gls{LPA}. In \gls{LPA}, the comparability condition can be viewed as a normalization condition for regulator functions. The order relation becomes independent of the \gls{RG} flow in this approximation but it depends on the number of spacetime dimensions. Within this analysis, we also provided an intuitive picture of how the behavior of \gls{RG} flows close to the aforementioned singular points is related to properties of the threshold functions.

For \gls{RG} flows in \gls{LPA}, we find that the Litim regulator provides the strongest singular behavior and is therefore the unique optimal regulator according to \gls{POSS}, at least within the standard set of regulator functions considered in the context of functional \gls{RG} studies. This implies that, in \gls{LPA}, the well-known functional optimization~\cite{Pawlowski:2005xe} and \gls{POSS} single out the same regulator as optimal: That is the Litim regulator. Moreover, we constructed a regulator which is close to the Litim regulator with respect to \gls{POSS} but is differentiable. In general, our analysis clearly indicates that analytic regulators generate significantly less stable \gls{RG} flows than the Litim regulator. In addition, by extending the set of functions associated with the standard regulators, we found that we can even construct regulators which come with a stronger singular behavior than the one exhibited by the Litim regulator and therefore generate particularly stable \gls{RG} flows in the spirit of~\gls{POSS}.

We close by emphasizing that \gls{POSS} does {\it a priori} not entail a minimization of the regulator dependence of physical observables. In other words, it is not apparent that functional optimization and \gls{POSS} in general single out the same regulator(s). Therefore, a natural extension of our analysis represents the computation of, e.g., critical exponents with regulators considered to be {\it good} according to \gls{POSS}, in particular with those characterized by a particularly strong singular behavior. Moreover, an analysis of how fast physical observables converge in the \gls{RG} flow may be interesting as this should be solely encoded in the dependence of the threshold functions on the mass parameter which is determined by the regulator, see \cref{fig:regulator_hierarchy}. In any case, given our analysis in \gls{LPA}, where the Litim regulator is singled out by both optimization criteria (within the standard set of regulator functions), it appears at least reasonable to expect that, beyond \gls{LPA}, the application of both criteria may also yield a set of regulator functions which may be simultaneously classified as ``good regulators", i.e., regulators which yield stable \gls{RG} flows according to \gls{POSS} and are associated with a ``short" trajectory in theory space according to the definition in Ref.~\cite{Pawlowski:2005xe}. Smooth regulators, which are singled out in \gls{LPA} by \gls{POSS}, may represent a good starting point for the construction of such regulators for computations of the effective action at higher orders of the derivative expansion.

\acknowledgements
We thank A.~Gei\ss el, A.~K\"onigstein, J.~M.~Pawlowski, M.~J.~Steil, and N.~Wink for useful discussions and comments on the manuscript. J.B. acknowledges support by the DFG under grant BR~4005/4-1 and BR~4005/6-1 (Heisenberg program). As members of the fQCD collaboration~\cite{fQCD}, the authors also would like to thank the other members of this collaboration for discussions. Moreover, this work is supported by the \textit{Deutsche Forschungsgemeinschaft} (DFG, German Research Foundation) through the CRC-TR 211 ``Strong-interaction matter under extreme conditions'' -- project number 315477589 -- TRR 211 and by the State of Hesse within the Research Cluster ELEMENTS (Project No. 500/10.006). All numerical results as well as all figures in this work have been obtained by using \textit{Python 3}~\cite{10.5555/1593511} with various libraries~\cite{2020SciPy-NMeth,Hunter:2007,harris2020array}, if not explicitly stated otherwise. Some of the results have been cross-checked with \texttt{Mathematica}~\cite{Mathematica:12.1}.
\appendix
\section{Constraints for regulator functions}\label{app:threshold-function-convergence}
For the following discussion of constraints for the regulator, we focus on threshold functions as they appear in \gls{LPA}. It then follows that a general regulator shape function $r$ should render the threshold functions appearing in the flow equations finite for $\omega > -\min_y P^2(y)$. Moreover, the threshold functions should diverge at $\omega = -\min_y P^2(y)$, see our discussion in Subsec.~\ref{sec:poss_LPA}. In general, the standard constraints detailed in \cref{eq:regulator-shape-function-constraints} or, equivalently, in \cref{eq:P2-constraints} are not sufficient in this respect. Hence, it is necessary to identify additional constraints which supplement the standard constraints for regulator shape functions. As we shall discuss below, these additional constraints in general depend on, e.g., the number of spacetime dimensions $d$ and on the position of the poles of the integrand associated with the threshold functions.

In the following we shall assume that the regulator shape function and its derivative behave as $r(y)\sim y^{-\alpha}$ and~$r'(y) \sim y^{-\alpha - 1}$ in the limits~$y\to 0$ and~$y\to\infty$, respectively. Constraints for the parameter~$\alpha$ are derived below. For concreteness, we use the normalization \eqref{eq:normalization-condition}, i.e., $\min_y P^2(y) = 1$. However, the following analysis does not rely on this normalization. 

Let us begin by discussing the issue of finiteness of the threshold functions, i.e., we first consider $\omega > -1$. In the limit $y \to \infty$, the integrand~\eqref{eq:generic-threshold-function-integrand} of the threshold function~\eqref{eq:generic-threshold-function} then becomes independent of~$\omega$ since $P^2(y)$ dominates the denominator: 
\begin{align*}
	\numberthis
	I^{(d)}_m(\omega, \dots) \sim y^{d/2+1} \frac{y^{-\alpha-1}}{y^m} \; \text{for} \; y \to \infty \, .
\end{align*}
Consequently, for $y \gg 1$, it is required that
\begin{align*}
	\numberthis
	-\alpha < -1 - \frac{d}{2} + m
\end{align*} 
to render the threshold functions finite. Recall that the parameter~$m$ counts the number of internal lines of the 1PI diagram associated with the threshold function~\eqref{eq:generic-threshold-function}.

In the limit $y \to 0$ and $\omega > -1$, we have to distinguish between two cases:
\begin{itemize}
	\item[(i)] 	$\lim_{y\to 0} P^2(y) < \infty$, i.e., $\alpha = 1$. For $\omega > -1$, we find
	\begin{align*}
		\numberthis\label{eq:y_to zero_alpha=1}
		I^{(d)}_m \sim \, c_m(y,\omega) y^{d/2+1} y^{-\alpha-1} \; \text{for} \; y \to 0 \, ,
	\end{align*}
	where~$c_m(y,\omega)$ depends on~$\omega$ and on~$y$ but is finite for~$\omega > -1$: 
	\begin{align*}
		\numberthis
		c_m(y,\omega) = \frac{1}{(P^2(y)+\omega)^m}\,.
	\end{align*}
	Finiteness of the threshold function only requires $d > 0$. Note that the integrand $I^{(d)}_m$ is nevertheless divergent for $d < 2$, see also~\cref{fig:threshold_function_integrand}.
	\item[(ii)] 	$\lim_{y\to 0} P^2(y) = \infty$, i.e., $\alpha > 1$. In this case, $\omega$ can be ignored, i.e., 
	\begin{align*}
		\numberthis
		P^2(y) + \omega \approx P^2(y)
	\end{align*}
	for $y\to0$. We then find
	\begin{align*}
		\numberthis
		I^{(d)}_m \sim y^{d/2+1} \frac{y^{-\alpha-1}}{y^{m-\alpha m}} \; \text{for} \; y \to 0 \, .
	\end{align*}
	Consequently, it is required that
	\begin{align*}
		\numberthis
		- \alpha > -1 -d/2 + m - m \alpha \, 
	\end{align*}
	to render the threshold functions finite. Note that, in this case, $y_0$ assumes a finite value. Recall that~$y_0$ is defined by~$P^2(y=y_0)=1$, see also the discussion of \cref{eq:introofy0}.
\end{itemize}

Next, we turn to an analysis of the behavior of the threshold functions for $\omega \to -1$. In this limit, the threshold functions exhibit a singular behavior, if $m \geq 1/2$ and if the pole of the integrand is located at a finite value~$y_0$, i.e., $y_0 > 0$ with $P^2(y=y_0) = 1$. In the following this is illustrated for analytic regulator shape functions with only one nonzero $y_0$. With \cref{eq:estimate1,eq:estimate2,eq:estimate3}, we find for such regulators that
\begin{align*}
	\numberthis
	I^{(d)}_m(\omega, r(y), r'(y), y) &>  -y^{\frac{d}{2}+1} \frac{\partial_y r_{\mathrm{Litim}}(y)}{(P^2(y) + \omega)^m} \\
	&\quad = \frac{y^{\frac{d}{2}-1}}{(P^2(y) + \omega)^m} \, 
\end{align*}
for~$y<y_0$. By performing now a Taylor expansion of $P^2(y)$ about $y_0$, we find for some fixed $\delta > 0$ that the integral
\begin{align*}
	\int_{y_0 - \delta}^{y_0} \d y\, \frac{y^{\frac{d}{2}-1}}{\left(\frac{1}{2} \partial_y^2 P^2 (y=y_0)(y-y_0)^2 + 1+\omega\right)^m}
\end{align*}
diverges for $m \geq 1/2$ and $d>0$ in the limit $\omega \to -1$. Consequently, $l^{(d)}_m(\omega)$ diverges in this limit. However, if $y_0=0$ is the only contact point (i.e., $r(y) \sim 1/y$), we have $c_m(y,\omega) \sim 1/y^m$ in \cref{eq:y_to zero_alpha=1} at $\omega = -1$ for \mbox{$y \to 0$}. Hence, the integral does not diverge in the limit \mbox{$\omega \to -1$}, at least for $d/2 > m$. For example, the standard exponential regulator $r_{\mathrm{exp}}(y)=1/(\exp(y)-1)$ (see \cref{fig:inverse_propagator}) yields a global minimum of $P^2(y)$ at $y_0=0$ and therefore the threshold function $l^{(d)}_m$ with $m=1$ does not diverge at~$\omega = -1$ for $d > 2$ in this case, see also Ref.~\cite{Pelaez:2015nsa}. According to \gls{POSS} and our corresponding numerical analysis presented in the main text, this class of regulators in general generates (numerically) unstable \gls{RG} flows in situations where the ground state is governed by spontaneous symmetry breaking, at least in studies where the full field dependence of the potential is considered.

We close our analysis by noting that neither the finiteness of the threshold functions for $\omega > -1$ nor the presence of a divergence in the limit $\omega \to -1$ follow necessarily from the constraints~\eqref{eq:regulator-shape-function-constraints} for general regulator shape functions. For example, according to our analysis and \gls{POSS}, suitable general purpose regulators in all dimensions $d>0$ and for all $m\geq 1/2$ are those which diverge as $1/y$ for $y \to 0$, approach zero in the limit $y \to \infty$ faster than $y^{-d/2-1}$ and yield an integrand of the threshold functions with at least one nonzero pole position~$y_0$.

\vfill

\bibliography{bib}

\end{document}